\documentclass[a4paper]{spie}  %>>> use this instead for A4 paper
\usepackage{amsmath,amsfonts,amssymb}
\usepackage{graphicx}
\PassOptionsToPackage{hyphens}{url}\usepackage[hidelinks]{hyperref}
\usepackage{upgreek}
\usepackage{xcolor}
\usepackage{lineno}
\title{On-Sky Single-photon Time resolution of 35 ps with White Rabbit synchronization: towards the measurement of the size of a White Dwarf star}

\author[1,2]{Federico Izraelevitch-Patitucci}
\author[1]{Sarah Tolila}
\author[1]{Ilian Ellafi}
\author[3]{Jean-Pierre Rivet}
\author[1]{Mathilde Hugbart}
\author[1]{Guillaume Labeyrie}
\author[3]{Olivier Lai}
\author[1]{Andreas Zmija}
\author[1]{William Guerin}
\author[1]{Robin Kaiser}

\affil[1]{Université Côte d'Azur, CNRS, Institut de Physique de Nice, France.}
\affil[2]{Instituto de Ciencias Físicas (ICIFI, UNSAM - CONICET), Argentina.}
\affil[3]{Universite Cote d’Azur, Observatoire de la Cote d’Azur, CNRS, Laboratoire Lagrange, France.}

\authorinfo{Corresponding author: Federico Izraelevitch-Patitucci, federico.izraelevitch@univ-cotedazur.fr}

\begin{document} 
\maketitle

\begin{abstract}
The IC4Stars (Intensity Correlation for Stars) project aims to measure the diameter of the white dwarf star Sirius B, using Intensity Interferometry. In this work we present our latest efforts and the milestones achieved in the last year. We report laboratory characterization of the single-photon detectors, TDC and synchronization electronics. We describe an observation campaign where we demonstrated on-sky time resolution below~35~ps~RMS, synchronizing two TDCs using the White Rabbit protocol and a~30~m telecom fiber. We developed the data acquisition of the raw time tags, and an algorithm to compute the second-order correlation function.
\end{abstract}


\section{INTRODUCTION}
\label{secIntro}

The IC4Stars (Intensity Correlation for Stars) project aims to measure the diameter of a white dwarf star for the fist time, using intensity interferometry. In particular, we aim to measure the closest white dwarf, Sirius B, the companion star of Sirius A. According to theoretical predictions, it is expected that Sirius B has approximately the size of the Earth and the mass of the Sun, which yields an angular diameter of~$\sim 30\, \mathrm{\upmu as}$. To be able to resolve the star in the blue spectral range, a telescope separation of approximately one kilometer is needed. To be able to obtain a SNR of about 10 in a few nights of observation at 8-m-class telescopes~[\citenum{guerin2025stellar}], several technical challenges must be addressed. Among them, we can mention:

\begin{itemize}
  \item On-sky operation of single-photon detection with time resolution in the scale of tens of pico seconds.
  \item The synchronization of two instruments coupled to different telescopes, separated~$\sim$~1~km, with a precision better than the time resolution of the photon detectors.
  \item The recording of   the raw time tags measured by the instruments and the data management for on-line monitoring and analysis.
  \item The development of a computing-efficient algorithm to calculate the second-order correlation function, $g^{(2)}$, on-the-fly and in a posterior analysis.
  \item A wavelength multiplexing instrument, capable of separating the incoming star light in several wavelength channels, and to compute the second-order correlation function between wavelength-matching channels of the two instruments.
\end{itemize}

In this work, we report our efforts in the first four technical fronts: time-resolution single-photon detectors, synchronization and clock distribution between two instruments, and data management and computing. In Sec.~\ref{secLabTests} we describe characterization tests performed on the detectors and on the Data Acquisition System (DAQ), in their respective subsection. In Sec.~\ref{secOnSky} we describe our setup during an on-sky acquisition campaign at the Epsilon Telescope, C2PU (Observatoire de la Côte d'Azur, Nice) in March 2026. Finally, we present the main conclusions and our future perspectives.

%%%%%%%%%%%%%%%%%%%%%%%%%%%%%%%
%%%%%%%%%%%%%%%%%%%%%%%%%%%%%%%
\section{Laboratory characterization tests}
\label{secLabTests}

\subsection{Detector characterization}
\label{subsecDetectorCharac}

Our current instrument prototype is based on the PhotonPix single-photon detectors (Photonscore)~[\citenum{leopold2026c2pu}]. This photodetector is based on a photocathode to convert the incoming photons into photoelectrons, and a Microchannel Plate (MCP) as an electron multiplier. According to specifications, these detectors show jitters of some tens of pico seconds, Photon-Detection Efficiency (PDE) of~$\sim$~30~\% and a dark count rate below~100~Counts Per Second (CPS).

In the MCP section of the detector, a high voltage generates a cascade of collisions inside microscopic channels. The high voltage sets the electric field strength across the MCP stack, directly controlling the gain and the speed of the electron avalanche. As the voltage is increased, the electron multiplication becomes faster and more efficient, which improves timing performance because the output pulses become steeper and more uniform. We studied the jitter of the detector as a function of the Bias Voltage (that can be adjusted via an internal code in the provided software), in a setup described below.

One peculiar effect of these detectors is the local saturation. If one impinges a photon stream in a beam of small diameter, for example below~1~mm, a saturation occurs in the MCP at high fluxes. In this situation, the measured count rate is smaller than the number of impinging photons per unit time (times the PDE), and the jitter performance is degraded. Instead, if the same photon stream is broadened into a larger-diameter beam, illuminating the whole sensitive area, then the saturation occurs at a much higher photon flux and the measured count rate matches the number of photons per unit time in the stream (times the PDE).

\begin{figure} [ht]
   \begin{center}
   \begin{tabular}{c} %% tabular useful for creating an array of images
   \includegraphics[width=0.5\textwidth]{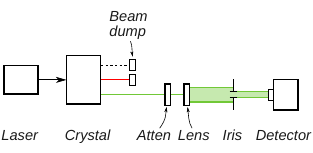}
   \end{tabular}
   \end{center}
   \caption[example]
%>>>> use \label inside caption to get Fig. number with \ref{}
   { \label{detectorCharacSchematics}
Characterization block diagram of the setup used to measure the Detector's  Jitter. A femtosecond-pulsed laser and a non-linear crystal are used to generate the probing beam.}
   \end{figure}

Figure~\ref{detectorCharacSchematics} shows the schematic diagram of the characterization experiments. The setup is based on a~1560~nm pulsed laser (Toptica) with a pulse width of~50~fs, an output power of~140~mW, and a repetition rate of~100~MHz. A nonlinear crystal, MgO:PPLN (Covesion), was used to generate the second and third harmonics, 780~nm and 520~nm, respectively. The latter is sufficiently close to our working wavelength of~417~nm, and was our operating wavelength. Once isolated, the~520~nm beam was attenuated and its diamter was controlled using a set of lenses and an iris. With the set of lenses, the beam diameter was broadened to ensure to have an uniform beam at the detector.

\begin{figure} [ht]
   \begin{center}
   \begin{tabular}{c} %% tabular useful for creating an array of images
   \includegraphics[width=1\textwidth]{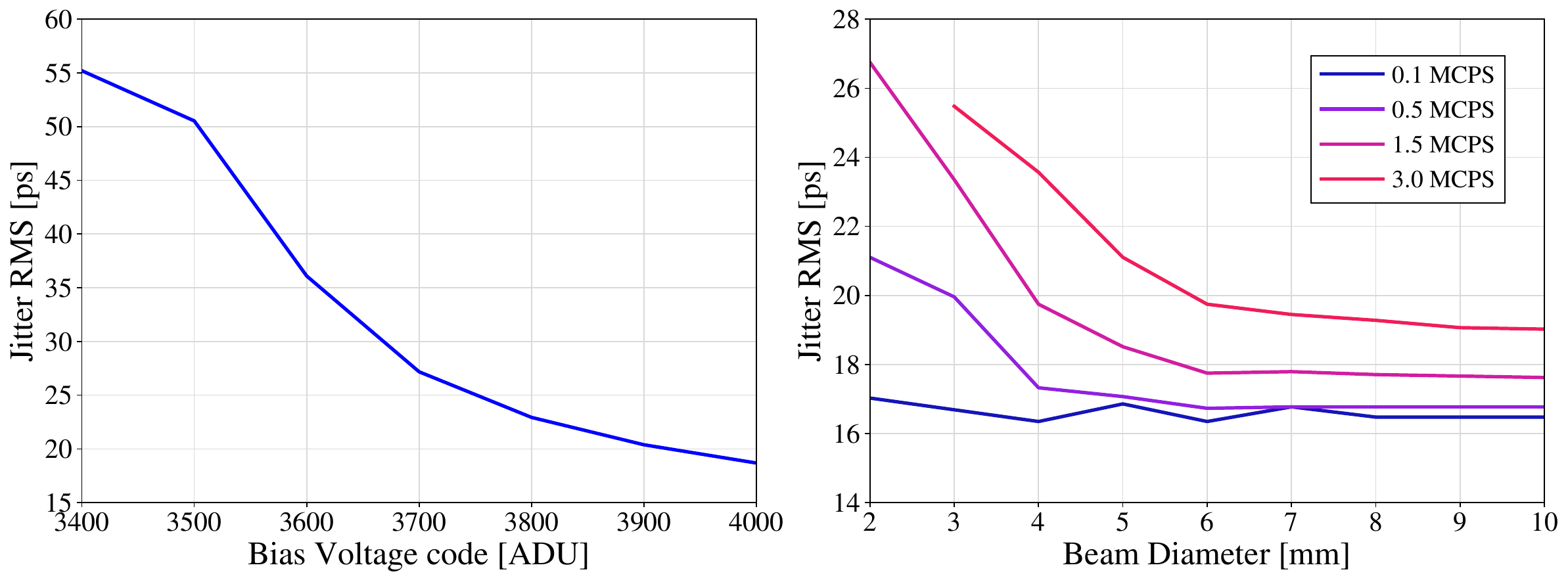}
   \end{tabular}
   \end{center}
   \caption[example]
%>>>> use \label inside caption to get Fig. number with \ref{}
   { \label{detectorJitterCharact_Proceeding}
Left: Detector Jitter as a function of the bias voltage, expressed in Analog-to-Digital Units, ADU (an internal code of the software that controls the detectors), using a~6-mm-diameter beam and performed at a~count rate of~$\sim 2$~MCPS. It can be seen that the Jitter decreases at higher Bias Voltage. Right: Jitter as a function of the Beam size, for different constant count rates. Curves acquired at maximum bias voltage and at a constant photon stream (constant number of photons per unit time per unit area). It can be seen that, when the beam is focused in a narrow spot, the Jitter increases due to the local saturation effect (see text for details). The~3.0~MCPS curve is truncated given that the count rate starts to decrease when the Beam diameter is smaller than~3~mm.}
   \end{figure}

In Figure~\ref{detectorJitterCharact_Proceeding}, we show two plots. The left-hand side one is the Jitter as a function of the detector internal high-voltage parameter (Bias Voltage code, expressed in Analog-to-Digital Units, ADU).  This measurement was performed using a collimated 6-mm-diameter laser beam at a count rate of approximately 2 MCPS. It can be seen that the Jitter decreases with an increasing voltage, and so we decided to operate at the highest bias voltage. The right-hand side plot shows the Jitter as a function of the impinging Beam Diameter, for different count rates, illustrating the local saturation effect previously mentioned. These curves were acquired at maximum bias voltage and at a constant photon stream (constant number of photons per unit time per unit area). As shown, when the beam is distributed in a larger area, the Jitter decreases. At high incident photon rates (above approximately 1.5 MCPS at large beam diameters) the detected count rate starts to drop towards small Beam Diameters. For example, when the curve of~3~MCPS reaches a beam diameter of~3~mm starting from a larger beam, the actual measured count rate is~1.9~MCPS (that is why we truncated the curve at this value). This characterization process was repeated for all the detectors used on-sky, and we report here a typical behavior.

\subsection{TDC and Synchronization characterization}
\label{subsecTDCSyncCharac}

The TDC used in our developments is the Time Tagger Ultra (Swabian Instruments), with a datasheet two-channels jitter below~11~ps. At our laboratory, we characterized the TDC itself using its internal clock, to measure the intrinsic jitter of the instrument. We used a Pulse Generator (Stanford Research Systems), square pulses, 50~\% of duty cycle, 2~V$_{\mathrm{pp}}$, zero offset, at~1~MHz. The instrument output was split with a passive Tee and connected to two inputs of the same TDC. We programmed the Pulse Generator to generate bursts of pulses every five minutes during a whole weekend, and we registered on disk the raw time tags of the two channels measured by the TDC. The obtained distribution of the time difference between the time tags of both inputs shows a Gaussian distribution. We performed a Gaussian fit to it and we obtained a sigma of~10.7~ps, consistent with specifications. This value varies slightly depending on the pair of channels used, and between TDCs. Here we display a typical response we obtained; the upper bound of this distribution was~$\sigma_{\mathrm{TDC}} = 11.0$~ps. This represents the contribution of the TDC to the overall time resolution of the system.

To synchronize two TDCs we used the White Rabbit protocol~[\citenum{WhiteRabbitTechnology}], implemented in two units of WR-LEN (Safran). These devices provide a~10~MHz clock and a~1~PPS signal that are connected to each TDC, disciplining the clocks of the TDCs. The internal clocks of each WR-LEN are synchronized by a telecom optical fiber and the aforementioned protocol. To characterize the intrinsic jitter of the WR-LENs, we connected the~10~MHz clock output of both units to two input channels of a~500-MHz oscilloscope. In this test, neither a Pulse Generator nor a TDC were used. We measured the signals provided by the WR-LENs directly connected to the oscilloscope. We performed the difference between the two rising edges and plotted the distribution. We obtained a Gaussian distribution with a sigma of~8.8~ps. This represents the contribution of the synchronization to the overall jitter of the system,~$\sigma_{\mathrm{WR}}$.

\begin{figure} [ht]
   \begin{center}
   \begin{tabular}{c} %% tabular useful for creating an array of images
   \includegraphics[width=0.6\textwidth]{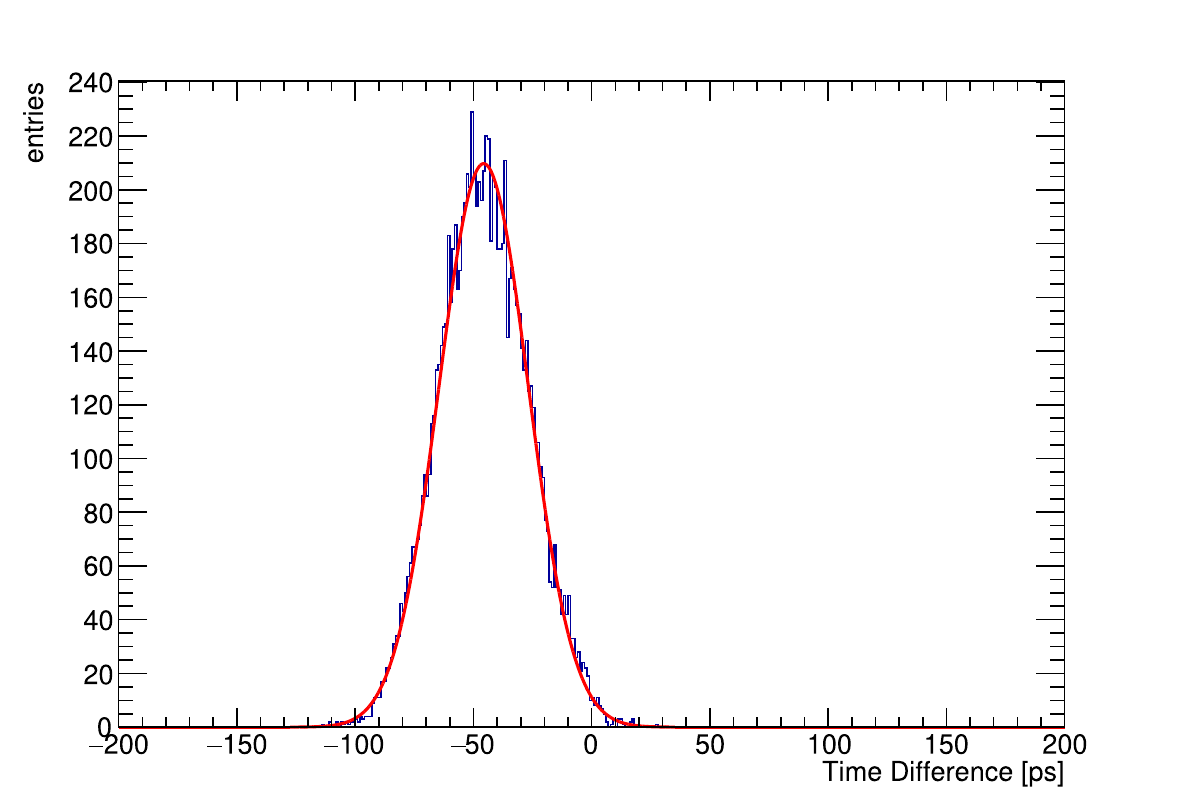}
   \end{tabular}
   \end{center}
   \caption[example]
%>>>> use \label inside caption to get Fig. number with \ref{}
   { \label{DAQ_atTheLab}
Distribution of the time difference between time tags registered by two independent TDC synchronized by the White Rabbit protocol, acquired at our laboratory. The inputs were fed by a square signal from a Pulse Generator. From the Gaussian fit we obtained $\sigma_\mathrm{DAQ}^{\mathrm{(lab)}}= 18.9$~ps~RMS, the total jitter of the DAQ at the laboratory.}
   \end{figure}

The complete system comprised the two TDCs and the two WR-LENs linked by a 30-m telecom fiber. Figure~\ref{DAQ_atTheLab} shows the time difference between time tags registered by the two independent TDCs and synchronized by two WR-LENs, obtained at our laboratory, for a 5-minutes lapse. The Pulse generator was used to provide the same excitation signal as previously described. From a Gaussian fit we obtained a $\sigma_\mathrm{DAQ}^{\mathrm{(lab)}}= 18.9$~ps~RMS. In this study we noticed that, if we restricted the acquisition to lapses below~10~s, then the obtained jitter was~$\sim 14$~ps. When considering longer acquisitions of real-case usage, then the mean of the distribution obtained in 10-s lapses drifted away. The result is that final distribution of long acquisitions is a broadened Gaussian with respect to the short acquisition case. At our laboratory, we obtained jitters below~19~ps for runs of one hour or more, which represented our benchmark for the contribution of the DAQ to the overall jitter.

%%%%%%%%%%%%%%%%%%%%%%%%%%%%%%%
%%%%%%%%%%%%%%%%%%%%%%%%%%%%%%%
\section{On-sky observation campaign}
\label{secOnSky}

\subsection{Experimental development}

\begin{figure}[ht]
   \begin{center}
   \begin{tabular}{c} %% tabular useful for creating an array of images
   \includegraphics[width=0.4\textwidth]{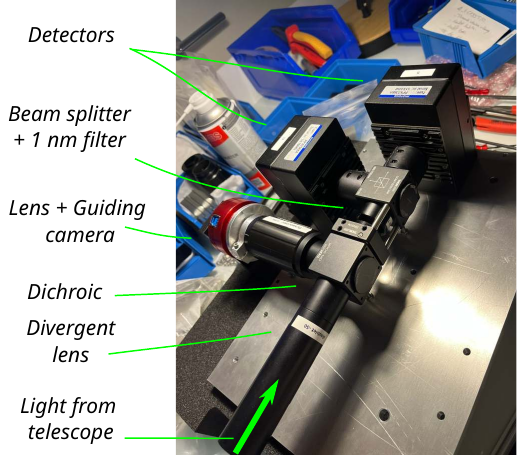}
   \end{tabular}
   \end{center}
   \caption[example]
%>>>> use \label inside caption to get Fig. number with \ref{}
   { \label{detectorModule}
Optical module, free-space coupled to the telescope. A Beam  Diameter of~6~mm was used to illuminate the Detectors.}
   \end{figure}

The experimental setup was deployed at the 1-m Epsilon telescope of C2PU (Centre Pédagogique Planète Univers), Observatoire de la Côte d’Azur, Plateau de Calern, in March 2026, following up our previous observation campaigns~[\citenum{rivet2020intensity, de2022combined, matthews2023intensity}]. It comprised a detector module, designed to be compact and easily transportable, and a data acquisition system (DAQ). The detector module is shown in Fig.~\ref{detectorModule}. The first optical element is a divergent lens, used to collimate the beam at the output of the telescope and to adjust its diameter to the optimal size for the experiment. We chose~6~mm as a compromise between timing performance and the minimization of the saturation effect (see right-hand side  plot of Fig.~\ref{detectorJitterCharact_Proceeding}, Sec.~\ref{subsecDetectorCharac}). Following the lens, a dichroic mirror was used to select (transmit) all wavelengths below~490~nm. The reflected beam was directed toward a guiding camera with its focusing lens. To ensure that the main beam is collimated onto the filter, the system was first aligned at the laboratory. We did so by verifying that the camera displayed a well-focused point, and then we fixed all the internal positions. Once the module was coupled to the telescope, the same focused point was observed on the camera as a verification. A bandpass filter was used (Alluxa), 417-nm central wavelength and~1-nm bandwidth. After the filter, a~50:50 non-polarizing beam splitter distributed the beam to both detectors.

\begin{figure} [ht]
   \begin{center}
   \begin{tabular}{c} %% tabular useful for creating an array of images
   \includegraphics[width=0.75\textwidth]{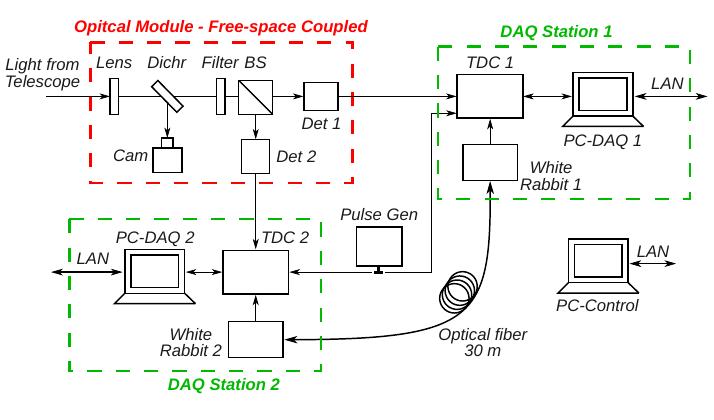}
   \end{tabular}
   \end{center}
   \caption[example]
%>>>> use \label inside caption to get Fig. number with \ref{}
   { \label{calernSchematicsNew}
Block diagram of the setup deployed on-sky. Each Detector of the Optical module was connected to a different DAQ Station, composed of a TDC, a White Rabbit module and a PC-DAQ. The latter are commanded via LAN by a Control PC located at the control room. The White Rabbit units were connected with a 30-m telecom fiber. A Pulse Generator was used to excite the TDCs and to characterize the DAQ system independently.}
   \end{figure}

Figure~\ref{calernSchematicsNew} shows a block diagram of DAQ. Each photon Detector was readout by a DAQ Station composed of a TDC, a WR-LEN and a PC. The internal clock of the TDC was disciplined by the WR-LEN. The two WR-LENs were connected through a telecom optical fiber of~30~m, providing a common clock to the TDCs. The Pulse Generator was connected to both TDCs using coaxial cables. As described in Sec.~\ref{subsecTDCSyncCharac}, the output of the Pulse Generator was split using a passive Tee and connected to inputs of each TDC. The PC of each DAQ Station stored the raw time tags locally. A Control PC, placed in the control room of the telescope building, was connected via Ethernet to the DAQ PCs. The Control PC commanded the start and stop of the acquisition runs, retrieved the raw time tags from the DAQ PCs, computed the second-order correlation function and produced plots that allowed us to visualize the experiment preliminary results on the spot.

\begin{figure} [ht]
   \begin{center}
   \begin{tabular}{c} %% tabular useful for creating an array of images
   \includegraphics[width=1\textwidth]{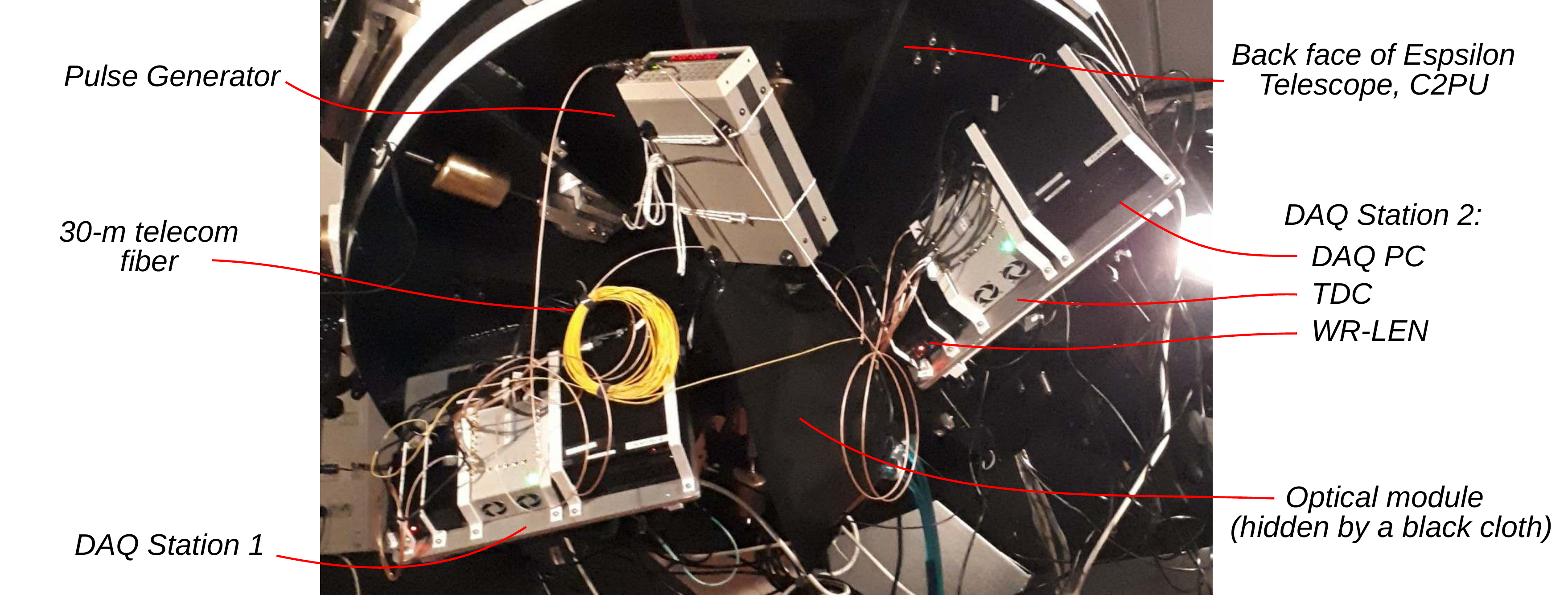}
   \end{tabular}
   \end{center}
   \caption[example]
%>>>> use \label inside caption to get Fig. number with \ref{}
   { \label{calernPicture}
Photograph of the setup coupled to the Epsilon Telescope, C2PU, Observatoire de la Côte d'Azur.}
   \end{figure}

Figure~\ref{calernPicture} shows a photograph of the back face of the telescope, where the setup was mounted. The optical module was mounted on the Cassegrain focus, and it was covered by a black cloth to prevent stray light to leak inside it (the black cloth makes difficult to identify the module in the picture). To minimize cable lengths, the instruments of each DAQ Station (TDC, WR and PC) were mounted onto a 12-mm aluminum plate and fixed to the back face of the telescope. Two of such Stations were mounted, one per Detector. The Pulse Generator was also mounted on the back face of the telescope.

We report here the observations during three nights, from 30-Mar-2026 through 1-Apr-2026. Given that our objective was to measure the bunching peak at zero baseline, i.e. the temporal second-order correlation function, we observed any bright-unresolved star. Considering their altitude during the night at that time of the year, we observed Regulus and Vega. The total observation time was~21.5~h and the average-weighted count rate was~1.6~MCPS. The observed count rate in both detectors was about~40~\% lower than the one expected, considering the stars luminosities, spectra, atmospheric extinction, optical throughput of the telescope and our instrument, and the PDE of the Detectors (as measured by the manufacturer at delivery time). This deficit in count rate is under investigation.

%%%%%%%%%%%%%%%%%%%%%%%%%%%%%%%
%%%%%%%%%%%%%%%%%%%%%%%%%%%%%%%
\subsection{On-sky Results}
\label{secResults}

\begin{figure} [ht]
   \begin{center}
   \begin{tabular}{c} %% tabular useful for creating an array of images
   \includegraphics[width=0.6\textwidth]{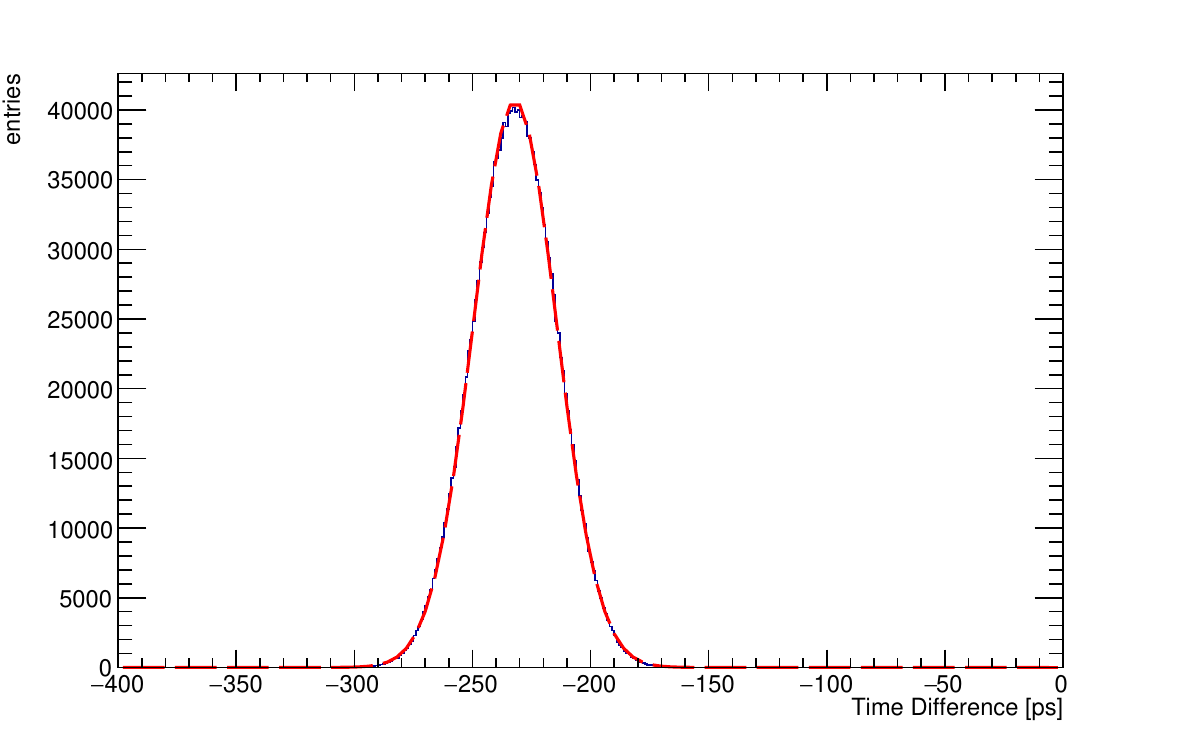}
   \end{tabular}
   \end{center}
   \caption[example]
%>>>> use \label inside caption to get Fig. number with \ref{}
   { \label{electronicJitter_Calern}
Distribution of the Time difference between time tags registered at the telescope the two independent TDC, synchronized by the White Rabbit protocol. The inputs were fed by a square signal from a Pulse Generator. The sigma obtained from the fit (red-dashed line) was~$\sigma_\mathrm{DAQ}^{\mathrm{(field)}} = 18.0$~ps and it represents the contribution of the DAQ system to the overall time resolution.}
   \end{figure}

Figure~\ref{electronicJitter_Calern} shows a plot of the time difference between the Pulse Generator signal measured by the two TDCs, synchronized by the WR-LENs, during a 30-minutes run. A Gaussian fit was performed and it is displayed over the histogram, in a red curve. The mean of the distribution is not at zero due to mm-scale cable-length differences. The sigma obtained from the fit (red-dashed line in the figure) was~18.0~ps. This represents the overall electronic jitter of the DAQ system obtained in the field experiment, $\sigma_\mathrm{DAQ}^{\mathrm{(field)}}$. This result is consistent with $\sigma_\mathrm{DAQ}^{\mathrm{(lab)}}$, and even slightly better that it. This discrepancy was expected, as we noticed systematic variations of~$\sim 1$~ps on this parameter every time we assembled the system.

\begin{figure} [ht]
   \begin{center}
   \begin{tabular}{c} %% tabular useful for creating an array of images
   \includegraphics[width=0.6\textwidth]{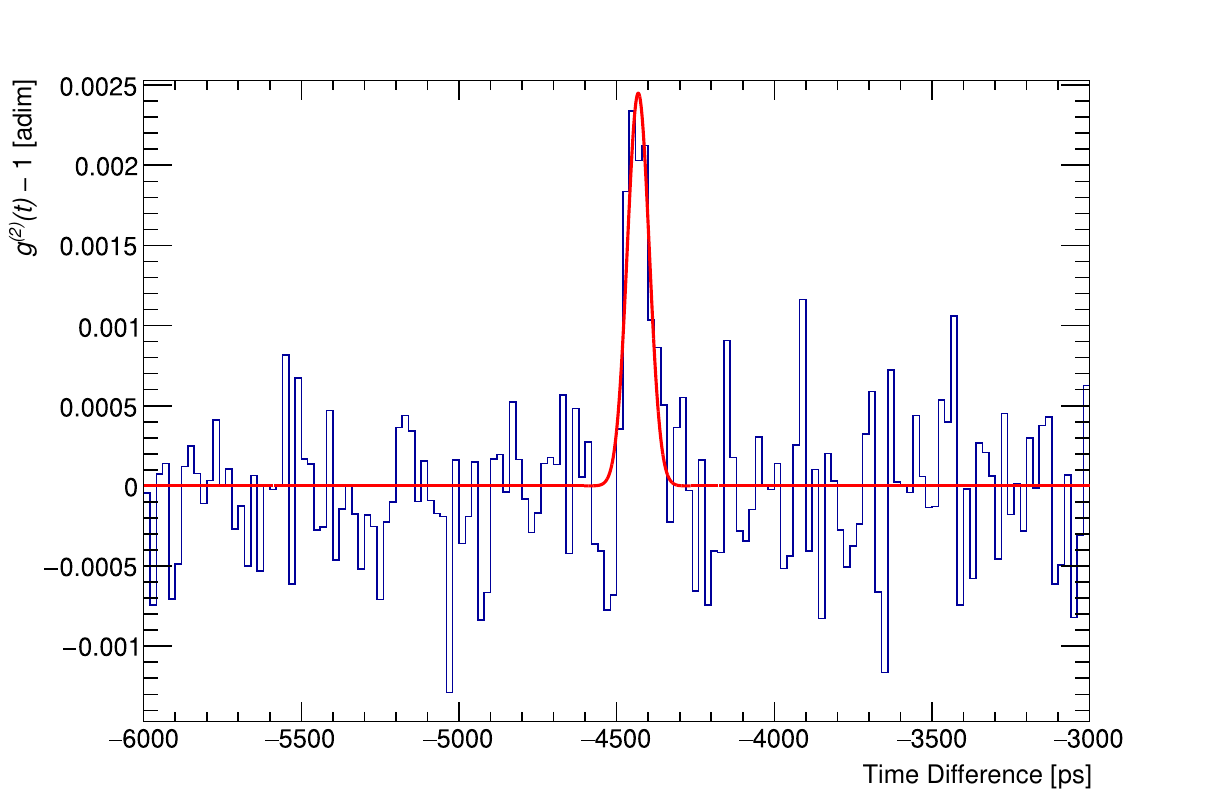}
   \end{tabular}
   \end{center}
   \caption[example]
%>>>> use \label inside caption to get Fig. number with \ref{}
   { \label{bunchingPeak}
Temporal second-order correlation function minus one, $g^{(2)}(t) - 1$, for the combined on-sky data set. From the Gaussian fit we obtained $\sigma_\mathrm{Tot}^{\mathrm{(field)}} = (34.7 \pm 4.4)$~ps (the overall time resolution obtained on-sky, including the DAQ and the two detectors), a coherence time of $(0.426 \pm 0.051)$~ps and an SNR~=~8.19. These results are consistent with the calculated expectations, within uncertainties.}
   \end{figure}

Figure~\ref{bunchingPeak} shows the temporal second-order correlation function minus one, $g^{(2)}(t) - 1$, for the combined data set. The bunching peak can be seen over a flat background due to uncorrelated detected photons. We performed a Gaussian fit from which we obtained the mean, the area, and the sigma. The mean obtained is consistent with the cable delay introduced, of $\sim 1$~m. The sigma from the fit was $\sigma_\mathrm{Tot}^{\mathrm{(field)}} = (34.7 \pm 4.4)$~ps, and it represents of the overall time resolution obtained on-sky, including the DAQ and the two detectors,

  \[ \sigma_\mathrm{Tot} =  \sqrt{\sigma_{\mathrm{DAQ}}^2 + 2 \sigma_{\mathrm{Det}}^2 } \;\;. \]

\noindent Inserting $\sigma_\mathrm{DAQ}^{\mathrm{(lab)}}$ and the detector time resolution determined at the laboratory, evaluated at the count rate obtained in the field (Sec.~\ref{secLabTests}), we obtained $\sigma_\mathrm{Tot}^{\mathrm{(lab)}} = 31.7$~ps, consistent with the field experiment within uncertainties.

The area obtained from the fit was $(0.213 \pm 0.026)$~ps, which yields an on-sky measured coherence time of $(0.426 \pm 0.051)$~ps, given that we measured unpolarized light. This measurement is consistent with the expected coherence Time of 0.492~ps, as calculated from the spectral filter transmission data~[\citenum{guerin2025stellar}]. The obtained SNR was~8.19, consistent with the expected SNR~of~8.28. The latter was obtained by computing the count-rate-based formula in Refs.~[\citenum{guerin2025stellar}]~and~[\citenum{dalal2024probing}], considering the three-parameters Gaussian fit.

%%%%%%%%%%%%%%%%%%%%%%%%%%%%%%%
%%%%%%%%%%%%%%%%%%%%%%%%%%%%%%%
\section{Conclusions and Outlook}
\label{secConclusions}

In this work we presented the prototype of our new instrument and the characterization of its components. We reported its deployment at the Epsilon Telescope, C2PU, Observatoire de la Côte d'Azur, on March 2026. We achieved the following milestones, towards the measurement of the angular diameter of Sirius B:

\begin{itemize}
  \item On-sky time-resolution of~$(34.7 \pm 4.4)$~ps~RMS of the overall instrument, including the contributions from the single-photon detectors, the TDCs and the synchronization system. This corresponds to a gain of a factor of~10 compared to our previous setups~[\citenum{guerin2025stellar}].
  \item The overall time resolution obtained on-sky was consistent, within uncertainties, with laboratory characterization measurements of the individual components.
  \item We synchronized on-sky two TDCs using the White Rabbit protocol and a telecom optical fiber of~30~m, with a time resolution of~18.0~ps~RMS.
  \item We implemented the measurement of the raw time tags of every detected photon by two different TDCs, stored in respective PCs.
  \item We developed and in-the-field implemented the second-order correlation function algorithm.
  \item The SNR and coherence time obtained on-sky resulted consistent with the calculations based on laboratory measurements. This result is crucial to predict longer observation campaigns and request telescope time.
\end{itemize}

Among the milestones to be addressed, we can mention:

\begin{itemize}
  \item The achieved time resolution implies that the position of the two telescopes shall be known at the~mm-scale, which represents a technical challenge.
  \item The observed count rate in the field was about~40~\% lower than the one expected, considering the full chain, from the starlight to photodetection. This deficit in count rate is currently under investigation. We plan to perform an independent measurement of the PDE of the detectors at our laboratory and to realize photometric studies using the Epsilon Telescope, to better understand this discrepancy.
\end{itemize}

In the near future, we will deploy the system at two separated telescopes, linked by the fiber, to perform spatial intensity interferometry using the synchronization scheme developed.

%%%%%%%%%%%%%%%%%%%%%%%%%%%%%%%
%%%%%%%%%%%%%%%%%%%%%%%%%%%%%%%
\section{Acknowledgments}
\label{secAcknowledgements}
We acknowledge the financial support of the French National Research Agency (project I2C, ANR-20-CE31-0003), the European project IC4Stars (ERC Advanced Grant No. 101140677).

% References
\bibliography{report} % bibliography data in report.bib

@article{rivet2020intensity,
  title={Intensity interferometry of {P Cygni in the H- $\alpha$ emission line: towards distance calibration of LBV supergiant stars}},
  author={Rivet, JP and Siciak, A and de Almeida, ESG and Vakili, F and Domiciano de Souza, A and Fouch{\'e}, M and Lai, O and Vernet, D and Kaiser, R and Guerin, W},
  journal={Monthly Notices of the Royal Astronomical Society},
  volume={494},
  number={1},
  pages={218--227},
  year={2020},
  publisher={Oxford University Press}
}

@article{de2022combined,
  title={Combined spectroscopy and intensity interferometry to determine the distances of the blue supergiants {P Cygni and Rigel}},
  author={de Almeida, ESG and Hugbart, Mathilde and Domiciano de Souza, Armando and Rivet, Jean-Pierre and Vakili, Farrokh and Siciak, Antonin and Labeyrie, Guillaume and Garde, Olivier and Matthews, Nolan and Lai, Olivier and others},
  journal={Monthly Notices of the Royal Astronomical Society},
  volume={515},
  number={1},
  pages={1--12},
  year={2022},
  publisher={Oxford University Press}
}

@article{matthews2023intensity,
  title={{Intensity interferometry observations of the {H-$\alpha$} envelope of $\gamma$ cas with M{\'e}o and a portable telescope}},
  author={Matthews, Nolan and Rivet, Jean-Pierre and Vernet, David and Hugbart, Mathilde and Labeyrie, Guillaume and Kaiser, Robin and Chab{\'e}, Julien and Courde, Cl{\'e}ment and Lai, Olivier and Vakili, Farrokh and others},
  journal={The Astronomical Journal},
  volume={165},
  number={3},
  pages={117},
  year={2023},
  publisher={The American Astronomical Society}
}

@article{leopold2026c2pu,
  title={82-ps Single-Photon Stellar Intensity Interferometry of {Vega}},
  author={Leopold, Verena and others},
  journal={to be published},
  year={2026},
}

@misc{WhiteRabbitTechnology,
  author       = {{White Rabbit Collaboration}},
  title        = {{White Rabbit Technology}},
  howpublished = {\url{https://www.white-rabbit.tech/wr-technology/}},
  note         = {Accessed: 2026-06-05},
}

@article{dalal2024probing,
  title={Probing {H0} and resolving {AGN} disks with ultrafast photon counters},
  author={Dalal, Neal and Galanis, Marios and Gammie, Charles and Gralla, Samuel E and Murray, Norman},
  journal={Physical Review D},
  volume={109},
  number={12},
  pages={123029},
  year={2024},
  publisher={APS}
}

@article{guerin2025stellar,
  title={Stellar intensity interferometry in the photon-counting regime},
  author={Guerin, William and Hugbart, Mathilde and Tolila, Sarah and Matthews, Nolan and Lai, Olivier and Rivet, Jean-Pierre and Labeyrie, Guillaume and Kaiser, Robin},
  journal={Comptes Rendus. Physique},
  volume={26},
  number={G1},
  pages={659--679},
  year={2025}
}
\bibliographystyle{spiebib} % makes bibtex use spiebib.bst

\end{document}